\documentclass[aps, prb,
floatfix,twocolumn,superscriptaddress
]{revtex4-2}
\usepackage[bookmarksopen=true, bookmarksopenlevel=1]{hyperref}
\usepackage{graphicx}
\usepackage{amssymb}
\usepackage{amsmath}
\usepackage{dcolumn}
\usepackage{bm}
\usepackage{color}
\usepackage{float}
\usepackage{natbib}
\usepackage{footnote}
\usepackage[normalem]{ulem}
\usepackage{hhline}
\begin{document}

\title{Lifetime of quasi-particles in the nearly-free electron metal sodium}

\author{D.V.\,Potorochin}
\affiliation{Institute of Experimental Physics, TU Bergakademie Freiberg, Leipziger Straße 23, 09599 Freiberg, Germany}
\affiliation{European XFEL, Holzkoppel 4, 22869 Schenefeld, Germany}
\affiliation{Deutsches Elektronen-Synchrotron DESY, Notkestrasse 85, 22607 Hamburg, Germany}
\author{R.\,Kurleto}
\affiliation {Leibniz Institute for Solid State and Materials Research  Dresden, Helmholtzstr. 20, 01069 Dresden, Germany}
\affiliation {Department of Physics, University of Colorado, Boulder, CO, 80309, USA }
\author{ O.J.\,Clark}
\affiliation{ Helmholtz-Zentrum Berlin, Albert-Einstein-Strasse 15, 12489 Berlin, Germany}
\author{ E.D.L.\,Rienks}
\affiliation{ Helmholtz-Zentrum Berlin, Albert-Einstein-Strasse 15, 12489 Berlin, Germany}
\author{ J.\,S\'anchez-Barriga}
\affiliation{ Helmholtz-Zentrum Berlin, Albert-Einstein-Strasse 15, 12489 Berlin, Germany}
\author{F.\,Roth}
\affiliation{Institute of Experimental Physics, TU Bergakademie Freiberg, Leipziger Straße 23, 09599 Freiberg, Germany}
\affiliation{Center for Efficient High Temperature Processes and Materials Conversion (ZeHS), TU Bergakademie Freiberg, 09599 Freiberg, Germany}
\author{V.\,Voroshnin}
\affiliation{ Helmholtz-Zentrum Berlin, Albert-Einstein-Strasse 15, 12489 Berlin, Germany} 
\author{A.\,Fedorov}
\affiliation{Leibniz Institute for Solid State and Materials Research  Dresden, Helmholtzstr. 20, 01069 Dresden, Germany}
\author{W.\,Eberhardt}
\affiliation{Center for Free-Electron Laser Science/DESY, 22607 Hamburg, Germany}
\author{B.\,B\"uchner}
\affiliation{Leibniz Institute for Solid State and Materials Research  Dresden, Helmholtzstr. 20, 01069 Dresden, Germany}
\affiliation {Institut f\"ur Festk\"orperphysik,  Technische Universität Dresden, 01062 Dresden, Germany}
\author{J.\,Fink}
\email[Corresponding author. E-mail address: ] {J.Fink@ifw-dresden.de.}
\affiliation{Leibniz Institute for Solid State and Materials Research  Dresden, Helmholtzstr. 20, 01069 Dresden, Germany}
\affiliation {Institut f\"ur Festk\"orperphysik,  Technische Universität Dresden, 01062 Dresden, Germany}

\date{\today}

\begin{abstract}\
We report a high-resolution angle-resolved photoemission (ARPES)  study of the prototypical nearly-free-electron  metal sodium.  The observed mass enhancement is   slightly smaller than that derived in previous studies. The new results on the lifetime broadening  increase the demand for theories beyond the random phase approximation.  Our results do not support the proposed strong enhancement of the scattering rates of the charge carriers due to a coupling to spin fluctuations. Moreover, a comparison with earlier electron energy-loss data on sodium  yields a strong reduction of the mass enhancement  of dipolar electron-hole excitations compared to that of monopole hole excitations, measured by ARPES. 

\end{abstract}


\maketitle

\section{\label{sec:intro} I. INTRODUCTION} 
The understanding of electron-electron ($e-e$) many-body interactions in metals is an ongoing challenge in solid-state physics since many decades.
These interactions are of great interest because they determine transport, thermal, and  magnetic
properties of metals. There are also many indications, that $e-e$ interaction, e.g. spin fluctuations~\cite{Dahm2009}  mediates unconventional superconductivity in cuprates and ferropnictides. These interactions are the subject of numerous experimental and theoretical studies during the last decades. Very often, they start from the classical behavior of  normal metals (Fermi liquids) assuming that these are well understood. 
\par
Here we report  a study of the simplest nearly-free-electron metal Na.  Even in this prototypical Fermi liquid metal, there are unresolved issues. Numerous attempts have been made to understand the electronic structure including the  mass enhancement $m^*/m=1.28$, using integrated and angle-resolved photoemission spectroscopy (ARPES)~\cite{Damascelli2003,Sobota2021,Smith1969,Jensen1985,Plummer1987,Lyo1988}. 
The mass renormalization is related to the real part of the self-energy $\Re\Sigma$. Various state-of-the-art many-body techniques were used, starting with the random phase approximation (RPA)~\cite{Quinn1958}, theories beyond RPA including the GW formalism with all types of 
approximations (GW +)~\cite{Hopfield1965,Northrup1989,Dolado2001,Cazzaniga2012,Zhou2018,Mandal2021}. Correlation effects have been discussed in Refs.~\cite{Ng1986,Nilsson2017,Craco2019}. There are ongoing discussions about the influence of spin fluctuations~\cite{Zhu1986}. More recently, a strong increase of the scattering rate $\Gamma$ due to a strong coupling to spin fluctuations was predicted, but not a corresponding increase of the mass enhancement~\cite{Lischner2014}. $\Gamma$ is related to the inverse lifetime $\tau$ and to the imaginary part of the self-energy by the relation $\Gamma=-2Z\Im\Sigma$, where $Z$ is the renormalization function, which for a less correlated material is expected to be constant and close to the Fermi level equal to Z=m/m*~\cite{Lundquist1969}.
\par
The early ARPES studies on the mass enhancement of the quasi-particles in Na~\cite{Jensen1985,Plummer1987,Lyo1988}  were performed with an energy resolution of 0.3 eV. No linewidth analysis has been performed in these studies. Here we report  not only on the mass enhancement, but also on the energy dependence of the lifetime broadening with an energy resolution, which is improved by a factor of 10 compared to the earlier experiments. 
To the best of our knowledge, no data on the lifetime broadening of photoholes in Na have been published in the literature. 
\par 
Based on our new ARPES data we hope that we contribute to the general understanding of 
many-body problems in normal metals (Fermi liquids), which may also help to understand the normal state  of cuprate superconductors (marginal Fermi liquids~\cite{Varma1989}), and iron pnictides which show  ``super Planckian scattering rates``~\cite{Hartnoll2021,Fink2021,Hemmida2021}.
\par
\section{\label{sec:exp} II. EXPERIMENTAL} 
The experiments were performed at the $1^2$ ARPES endstation using the UE 112-PGM-2a beamline. Additional experiments were performed at the Spin-ARPES  end station of the U125-2-PGM beamline. As a substrate we used a W(110) single crystal. This crystal was cleaned before deposition by annealing in O atmosphere (1x$10^{-7} $ mbar) at  1200 $^\circ$C followed by flashing at 2200 $^\circ$C. The quality of the W(110) surface  was verified by LEED experiments, core level spectroscopy on the W $4f$ level, and by ARPES experiments. 

Na was evaporated out of a Mo crucible on the single crystalline W substrate, cooled to 20 K. Subsequently, the sample was annealed for 15 minutes at room temperature. Most of the results reported here were obtained from a Na(110) film having a thickness of about 15 nm. During measurements, the temperature of the crystal was 20 K. The base pressure of the experimental setup was better than $6\cdot 10^{-11}$ mbar.  Photoelectrons were detected with a Scienta R8000 electron analyzer. The overall experimental energy and momentum resolutions were set to 30 meV  and $0.2\,^{\circ}$, respectively.
\section{\label{sec:expres} III. EXPERIMENTAL RESULTS} 
In Fig.~1  we  show an energy-momentum distribution map measured with a photon energy of $h\nu=70$ eV along $k_{||}= k_y=\Gamma-N = <100>$  at a $k_z$ value corresponding to the $\Gamma$ point in the third  Brillouin zone (BZ). The $\Gamma$ point has been determined from similar photon-energy-dependent spectra described in Appendix A. We  determined the dispersion $E(k)$ by fitting momentum distribution curves (MDC) with Lorentzians. The white dashed line is a fit of the derived dispersion with a parabola. 
\par
\begin{figure}[t!]
\centering
\includegraphics[angle=0,width=1.0\linewidth]{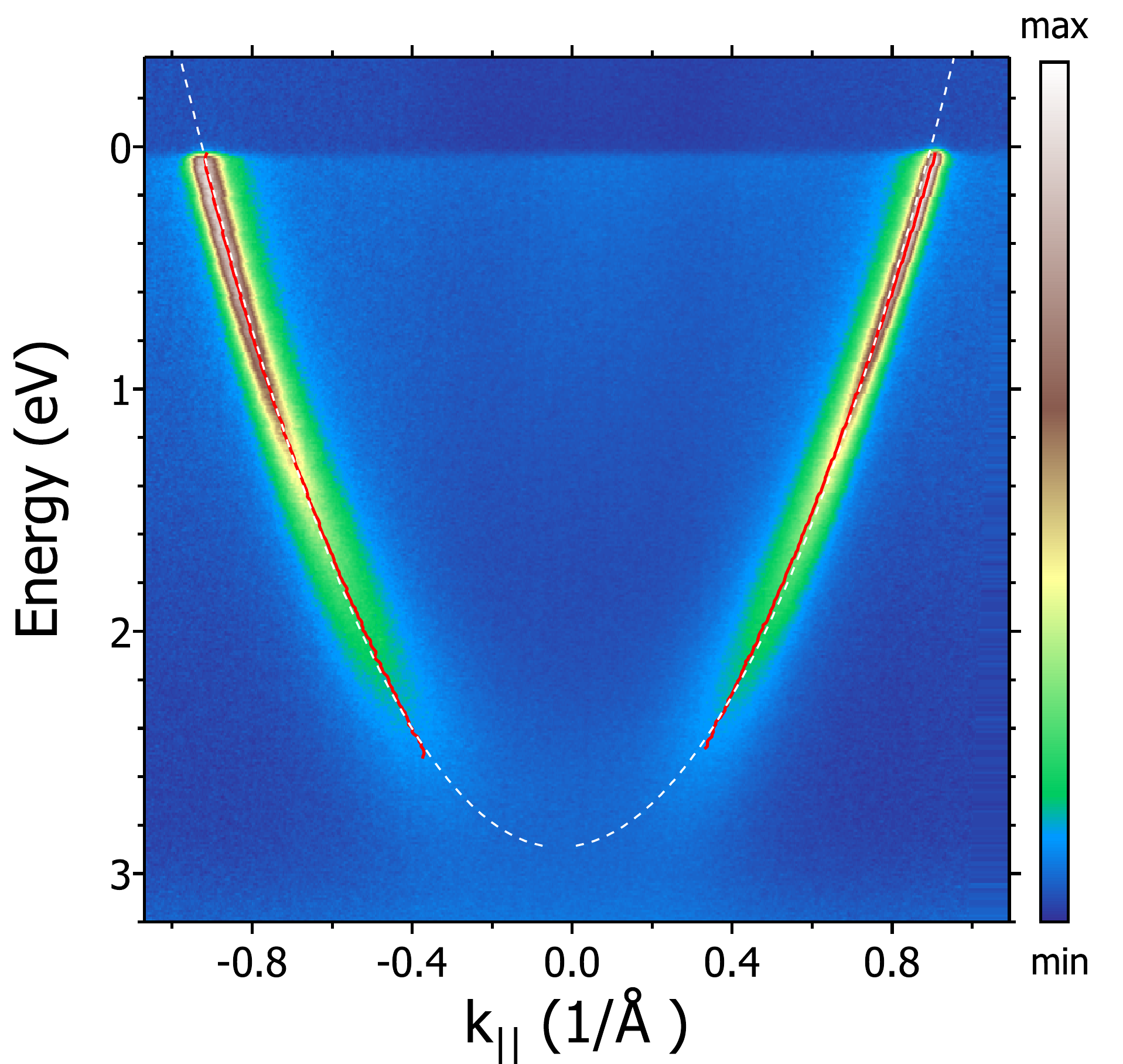}
 \caption
 {Experimental energy-momentum distribution map of Na measured along the $k_y$ direction using horizontally  polarized photons. Red line:  dispersion derived from a Lorentzian fit to momentum distribution curves. White dashed line: fit of the maxima of the Lorentzians with a parabolic dispersion.
}
\centering
\end{figure}
The spectral weight at the bottom of the band is rather low when compared with that at the Fermi level. The reason for this is related to matrix element effects. The initial state has predominantly $s$ character, i.e., it is even with respect to the $k_x-k_z$ mirror plane. We measured with photons having a horizontal polarization. This means the dipole operator is even relative to that mirror plane. 
The vanishing intensity could indicate that the final states, which according to dipole excitation must be of $p$-character, are predominantly odd with respect to the scattering plane.
The situation is very similar to our previous study of the waterfall dispersion of the spectral weight in Nd$_2$CuO$_4$~\cite{Rienks2014}.
\par
In Fig.~2 we depict the Fermi surface of  Na(110)   measured with a photon energy of $h\nu=70$ eV in the $k_z=\Gamma,k_x-k_y$ plane. The data were derived by a summation of intensities in an energy range of 0.015 eV close to the Fermi level. The Fermi surface is close to a circle as expected for a nearly-free-electron metal. The  Fermi wave vector along the $k_y$ line is $k_{\mathrm{F}}=0.91$ 1/{\AA}.
\begin{figure}[t!]
\centering
\includegraphics[angle=0,width=1.0\linewidth]{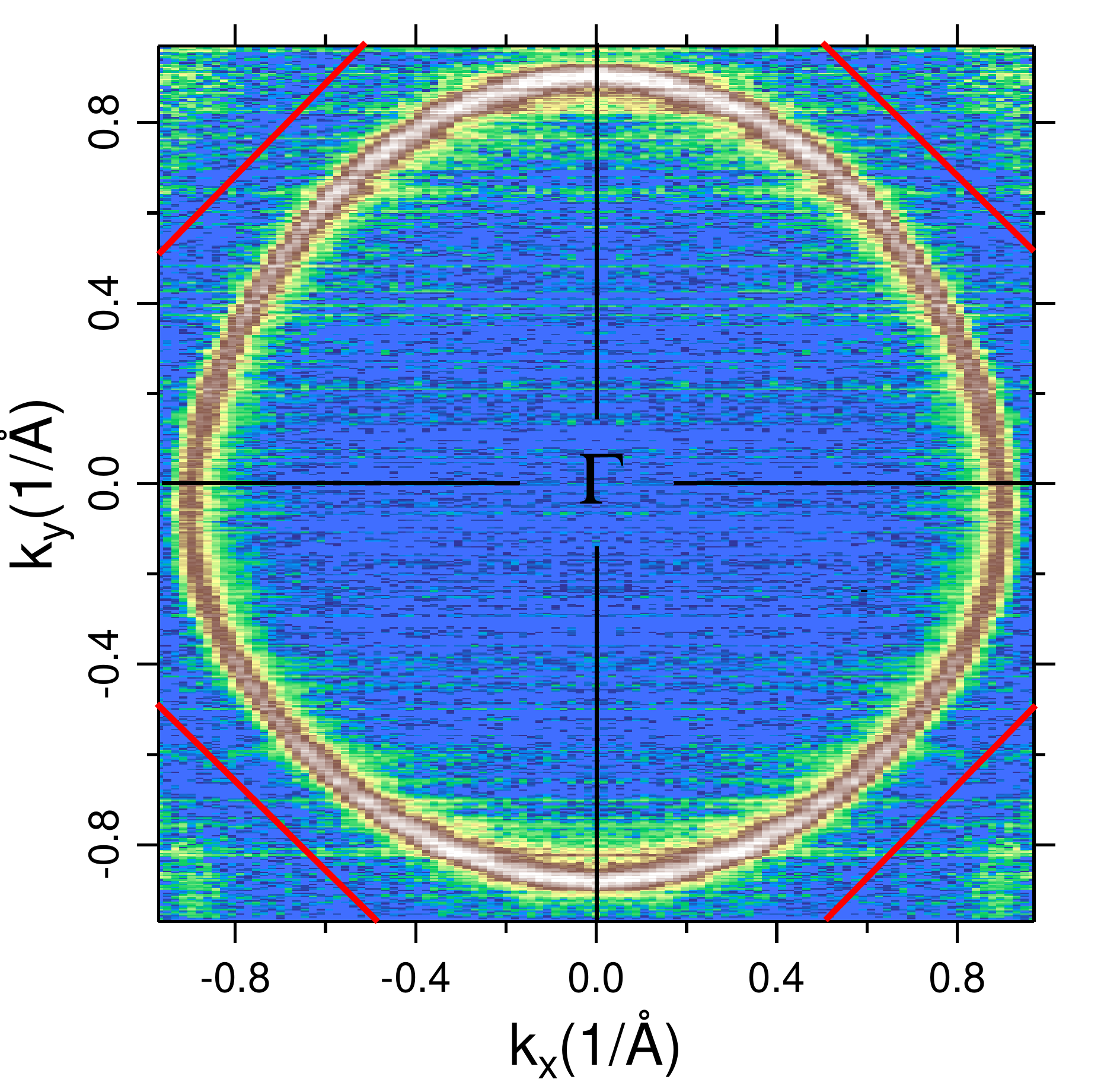}
\caption
 {Experimental Fermi surface  map of Na(110) measured with horizontally ($||~k_x$) polarized photons. The data are symmetrized relative to $k_x =0$. Red  lines: part of the Brillouin zone. 
}
\centering
\end{figure}
The small deviations from a circular Fermi surface detected in the present experiment are   caused possibly by a flattening and splitting of the dispersion due to the proximity to the Brillouin zone (BZ),  or by a non-perfect angle calibration of the lens parameters or alignment.
\par
In the corners of Fig.~2 we detect a faint signal of the Fermi surface of the second BZ. This observation indicates that the orientation of the Na(110) single crystal is along the main high symmetry lines. This also supports that  the momentum distribution map, presented in Fig.~1, is measured along the $<100>$ ($k_y$) direction. 
\begin{figure} [t!]
\centering
\includegraphics[angle=0,width=1\linewidth]{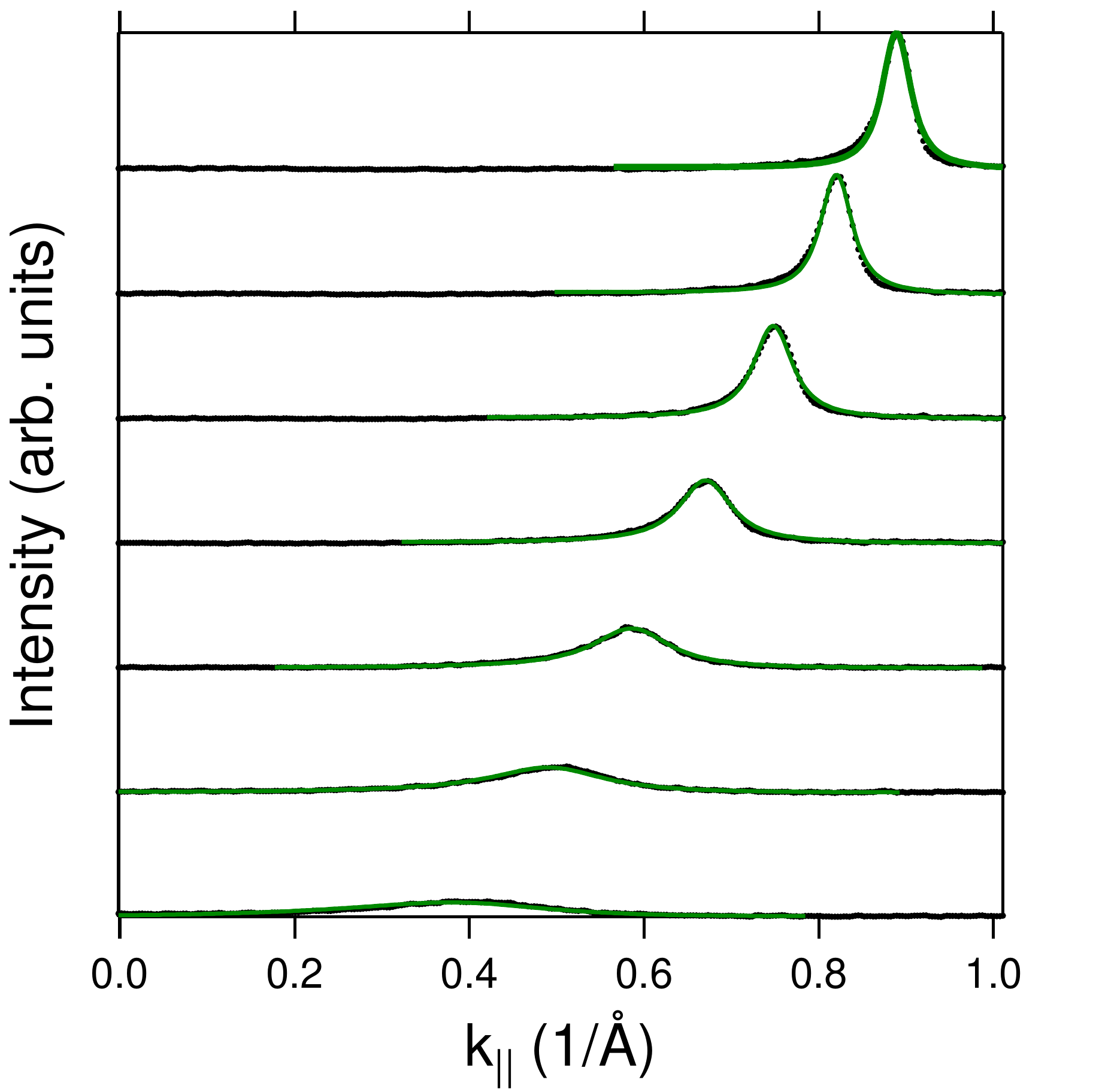}
\caption {Waterfall plot of ARPES momentum distribution curves of the valence band as a function of energy (black dots) in the energy range  $0.103 \leq E \leq 2.263$ eV, measured along $k_y$ (see Fig.~1 and 2).  Green line: ``all-at-once'' fits of the spectral function with energy dependent lifetime broadening and including corrections from finite energy and momentum resolution.} 
\centering
\end{figure}
\par 
In Fig.~3 we present a waterfall plot of momentum distribution curves as a function of energy together with an ``all at once fit''~\cite{Kurleto2021}. For the fit we use a quartic dispersion $E(k) =W - \gamma_1 k^2 - \gamma_2 k^4$. 
A small deviation from a free-electron parabolic dispersion was also derived from  band structure calculations for Na~\cite{Ching1975}. From the fit we obtain the following parameters: 
$W = $2.78 eV, $\gamma_1 =$3.77 eV{\AA}$^2$, $\gamma_2 =-0.49$ eV{\AA}$^4$, and the  total line width broadening 
$\Gamma(E)_{\mathrm{exp}}$, corrected for finite experimental resolution, for each energy. From a comparison of the bottom of the band $W$ with that derived from LDA band structure calculations ($W_{\mathrm{LDA}}=$3.18 eV~\cite{Ching1975}), a mass renormalization $m^*/m_0=$1.14  is calculated.
\par 
We emphasize, that we do not use the traditional method for the evaluation of the energy dependence of the line width, i.e., fitting the momentum distribution curves with Lorentzians and multiplication the derived widths in momentum space with the velocity. Rather we fit the two-dimensional data with the spectral function described with parameters related to the width in energy space at each energy. 
\par
To obtain the lifetime broadening of the photohole  $\Gamma^{\mathrm{in}}_{\mathrm{h}}(E)$ due to $e-e$ interaction, we have to subtract contributions from elastic scattering $\Gamma^{\mathrm{el}}_{\mathrm{h}}(E)$ and/or a broadening due to a finite inverse lifetime $\Gamma_{\mathrm{e}}$ of the photoelectrons. 
\par
First, we assume that the finite value $\Gamma^{\mathrm{el}}_{\mathrm{h}}(0)$  is completely determined by impurity scatterers. The constant mean distance $d$ between the impurities is equal to $1/\Delta k$, where $\Delta k$ is the momentum width at the Fermi level. Then the  broadening due to  elastic scattering  is given by $v(E)\Delta k$, where $v(E)$ is the energy-dependent group velocity taken from the experiments.
\par
Second, we assume that the broadening at $k_{\mathrm{F}}$ is caused by the finite lifetime  broadening $\Gamma_{\mathrm{e}}$ of the final-state photoelectron, which is usually termed  ``final state effect``. For normal emission, the final state broadening  leads at the Fermi level  to a broadening $(v_{\mathrm{h}}(E)/v_{\mathrm{e}}(E))\Gamma_{\mathrm{e}}$~\cite{Pendry1978,Eastman1978,Thiry1979,Knapp1979,Chiang1980,Kevan1980,Grepstad1982,Bartynski1986,Smith1993}, where $v_{\mathrm{h}}(e)$ and $v_{\mathrm{e}}(E)$  are the photohole and photoelectron group velocities, respectively. Then the inelastic scattering rate of the photoholes is given by $\Gamma^{\mathrm{in}}_{\mathrm{h}}(E)=\Gamma_{\mathrm{exp}}(E)-(v_{\mathrm{h}}/v_{\mathrm{e}})\Gamma_{\mathrm{e}}$. For the more general corrections in the off-normal case, we use the register-line formalism~\cite{Himpsel1980,Fraxedas1990,McLean1994}  
which is  discussed in detail in Appendix B. 
We conclude that in the analyzed energy range we get a very similar inelastic scattering rate $\Gamma^{\mathrm{in}}_{\mathrm{h}}(E)$  regardless of whether we apply corrections caused by elastic scattering or by ``final state effect``. Using the mean value of such  corrections we obtain the inverse hole lifetime $\Gamma^{\mathrm{in}}_{\mathrm{h}}(E)$ which is depicted in Fig.~4. 
The result can be fitted using the relation $\Gamma(E)=\alpha E^n$ with $\alpha= 0.131\pm0.012$ and $n=1.98\pm0.08$  (see Fig.~4). In Appendix B we depict the uncorrected curve which, at the Fermi level has a finite value of 0.18 eV. 
\par
The real and the imaginary part of the self-energy are connected by the Kramers-Kronig transformation (KKT). This means that an enhancement of the scattering rate leads in tandem to an enhancement of the effective mass. Using  the renormalization constant $Z=m/m^*$=0.88 from the band width renormalization we have calculated  $\Im\Sigma(E)=-\Gamma(E)/2Z$. Upon performing the KKT of  $\Im\Sigma(E)$ we obtain $\Re\Sigma(E)$. From this,  the mass renormalization $m^*/m=1+\Re\Sigma(E)/E=1.18$ can be calculated. This value is close to the value obtained directly from the  band renormalization (see above and Table I). 
\begin{figure} [t!]
\centering
\includegraphics[angle=0,width=1\linewidth]{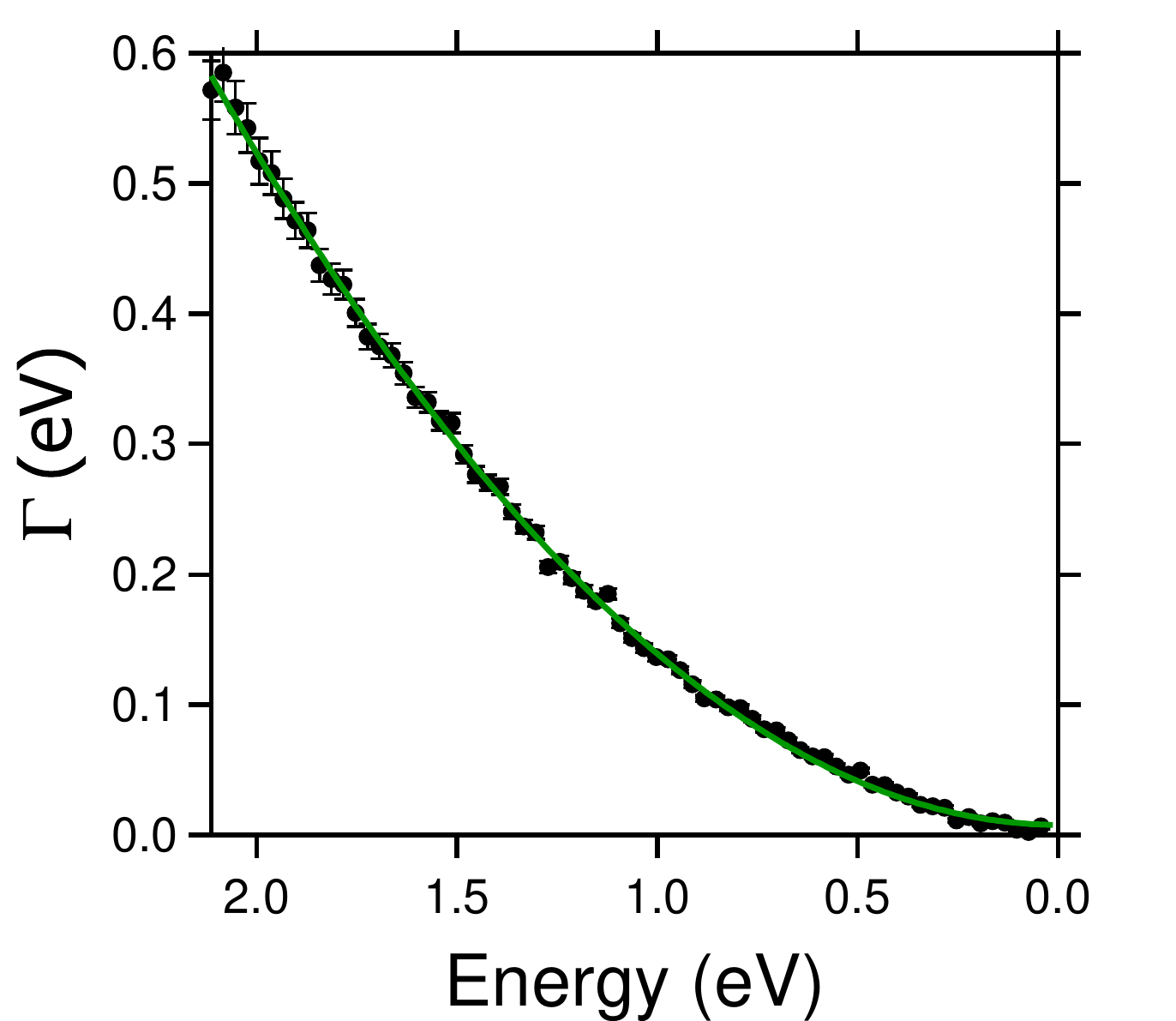}
\caption {Inelastic (e-e) scattering rates $\Gamma$ of photoholes in Na as a function of energy. Black dots: corrected ARPES data derived from the fit presented in Fig.~3. Green solid line: fit with  $\Gamma^{\mathrm{in}}_{\mathrm{h}}(E)=\alpha E^n$.} 
\centering
\end{figure}
\par
\section{\label{sec:disc} IV. DISCUSSION}
The dispersion, derived from the data in Fig.~1 is very close  to a parabola. This is expected for a nearly-free-electron metal. A possible kink due to electron-phonon coupling would appear at very low energies because of the low Debye energy $\Theta_D=0.013$ eV of Na~\cite{Martin1961} and it would be very weak because the electron-phonon coupling should be small in a nearly-free-electron system such as the non-superconducting Na. In Table~I we compare the band width and the mass enhancement (assuming a parabolic dispersion) with results from the Plummer group and theoretical calculations. The reduction of the band width when compared with that from LDA calculations~\cite{Ching1975} is slightly less than  that of the early ARPES study. Correspondingly, also the mass enhancement is slightly smaller. On the other hand, the deviations compared to the earlier values are not very pronounced. This means that the discussion of the band width reduction should follow earlier work. The free-electron RPA calculations yield mass enhancements which are too small and the GW approximation or local field corrections are needed to describe the experimental data (see Table I).
\par
The energy dependence of the line width is almost perfectly quadratic. The  exponent $n=$1.98$\pm$0.08 is very close to two. The quadratic energy dependence extends over a large energy range up to 2.2 eV corresponding to a temperature of 25000 K.
\par 
The prefactor $\alpha$ is considerably higher than that of the RPA free-electron value of Quinn and Ferrel ($\alpha=0.076$)~\cite{Quinn1958}.  Because the real part of the self-energy, determining the band width, and the imaginary part, determining the lifetime,  is connected via the KKT, this difference is expected when regarding the data in Table I. Thus, our data on the lifetime broadening demand that one has to go \textit{beyond} RPA. This is expected because Na with an $r_S$ value close to 4 is not a high-density Fermi liquid. Theoretical calculations using the GW approximation or local field corrections are closer to the experimental data. On the other hand, the  calculations by Lischner \textit{et al.}~\cite{Lischner2014} postulating the importance of spin fluctuations for the scattering rate yield a much too high pre-factor $\alpha$ (see Table I). Thus our experimental results do not support the proposed strong coupling  of the conduction electrons to spin fluctuations in Na.
\par
Comparing the $\alpha$ value  of the valence electrons in Na with  $\alpha=0.24$ derived for surface states in the $4d$ metal Mo~\cite{Valla1999} signals the expected enhanced scattering rate in $4d$ metals relative to that of $sp$ metals.
\par 
\begin{table}[t]
\centering
\caption{ARPES data of Na compared with theoretical data from the literature: ARPES,  present work; ARPES/KKT  effective mass derived by a Kramers-Kronig transformation of the scattering rate; ARPES  from Plummer's group; RPA FEG, free-electron gas; RPA LDA; GW LDA + plus approximations; GW LDA SF plus spin fluctuations. $W$: band width, $m^*/m$ mass renormalization using $W_{\mathrm{LDA}}=$ 3.18 eV from LDA calculations~\cite{Ching1975}. $\alpha$: pre-factor from the fit of the corrected scattering rates.  $\Gamma_{\mathrm{e}}$: lifetime broadening of the photoelectron} 
\label{tab:1}   
\begin{ruledtabular}
\begin{tabular}{ l l l l l  }
&    $W$ (eV)  & $m^*/m$ & $\alpha$ (1/eV) & $\Gamma_{\mathrm{e}}$ (eV)    \\ 
\cline{1-5}
ARPES \footnotemark[1]  & 2.78      & 1.14  &  0.131$\pm $0.012     &  $\lessapprox $4.8   \\ 
ARPES/KKT\footnotemark[2] &      & 1.18  &     &    \\ 
ARPES\footnotemark[3]  & 2.5-2.61  & 1.28 &            &         \\
RPA FEG\footnotemark[4]  & 2.96          & 1.10      & 0.05-0.077 &  3.3    \\
RPA LDA\footnotemark[5]  &           &      & 0.1        &  3.3    \\
GW + \footnotemark[6]& 2.52-2.89      & 1.11-1.26 & 0.10-0.34       &         \\
GW LDA SF\footnotemark[7] & 2.51      & 1.27 & 0.47       &         \\
\end{tabular}
\end{ruledtabular}
\footnotetext[1]{ARPES, present work}
\footnotetext[2]{ARPES+KKT, present work}
\footnotetext[3] {Refs.~\cite{Jensen1985,Plummer1987,Lyo1988}}
\footnotetext[4] {Refs.~\cite{Quinn1958,Hedin1965,Hopfield1965,Lundquist1969,Dolado2001,Chulkov2006}}
\footnotetext[5]{Ref.~\cite{Dolado2001}}
\footnotetext[6]{Refs.~\cite{Cazzaniga2012,Lischner2014}}
\footnotetext[7]{Ref.~\cite{Lischner2014}}
\end{table}
\par
Next we discuss a comparison of our ARPES data with data of   interband  and intraband excitations in Na~\cite{Felde1989} measured by EELS. From the cutoff of interband transitions a mass enhancement  $m^*/m =1.05 \pm 0.04$ was derived. A similar value $m^*/m $=1.0 was obtained from the analysis of zone boundary collective states (ZBCS),  a combination of intraband and interband excitations from the band bottom to states near the BZ.  The inconsistency between the ARPES and the EELS data on the effective mass was already previously noticed~\cite{Sturm1989}. It can be rationalized in the following way: EELS measures inter- or intraband excitations, i.e., two particle or electron-hole excitations with dipole character. They are  less screened than single-hole or monopole excitations detected in ARPES experiments.  
\par
Final state effects may have far-reaching implications for the interpretation of all ARPES data. E. g., in Ref.~\cite{Yasuhara1999} it is claimed, that the mass renormalization of the valence electrons detected in ARPES is related to final state effects. This interpretation can be ruled out because a similar mass enhancement  $m^*/m=1.256$ has been  detected in de-Haas-van-Alphen effect measurements~\cite{Elliott1982}.
\par
Assuming that the momentum width $\Delta k_{\mathrm{F}}=0.045 $ 1/{\AA} is predominantly caused by elastic scattering we derive  a mean-free path of the photoelectrons at the Fermi level of $1/\Delta k_{\mathrm{F}}=$22 {\AA}.
\par
Assuming that the broadening at the Fermi level is predominantly caused by the finite lifetime of the photoelectrons, we derive the result $\Gamma_{\mathrm{e}}=$4.8 eV. This value is in rather good agreement with  the value of about 3 eV estimated for 100 eV electrons in Na from an analysis of LEED data~\cite{Pendry1978} or the theoretical value calculated on the basis of a coupling to plasmons in a free-electron model (RPA)~\cite{Lundquist1969} or using RPA LDA~\cite{Dolado2001} predicting $\Gamma_{\mathrm{e}}$=3.3 eV. Using a free-electron approximation  one can also estimate the mean-free path of 70 eV photoelectrons in Na: $v_{\mathrm{e}}/\Gamma_{\mathrm{e}}=$ 6 {\AA}. This value  is close to 8~{\AA} resulting from core level photoelectron experiments~\cite{Smith1993}. The agreement of our results for the lifetime broadening and the mean-free path of the photoelectrons with those in the literature supports our evaluation of the lifetime of the photohole as a function of energy. 

\section{\label{sec:exp} V. SUMMARY}

\par
In the present high-resolution ARPES study of the prototypical nearly-free electron metal sodium, we  confirm previous results of the mass enhancement of the charge carriers, although with a slightly  smaller value. In addition, we present the energy dependence of the lifetime broadening. A perfect quadratic energy dependence is observed. The prefactor is enhanced compared to RPA calculations. The reason is that sodium is not  in the high-density region, thus expecting slightly larger scattering rates caused by many-body interactions. The central result of our sodium ARPES study  is, that it does not support the theoretical prediction of a rather strong enhancement of the lifetime broadening of the quasi-particles due to a coupling to spin fluctuations. A comparison with  excitations studied with EELS indicates, that those dipolar two-particle excitations are much less screened than the single-particle excitations studied by ARPES. 
 
\section{\label{sec:intro} ACKNOWLIDGMENTS}
J. S.-B. gratefully acknowledges financial support from the Impuls- und Vernetzungsfonds
der Helmholtz-Gemeinschaft.

\appendix 

\section{3D NATURE OF THE FERMI SURFACE}
In addition to the $k_{||}$ dependence of the Fermi wave vector presented in the main paper, we have also performed photon energy dependent ARPES experiments to obtain information on the $k_z$ dependence of the Fermi surface. ARPES data of the Fermi wave vector $k_{\mathrm{F}}$  as a function of the wave vector $k_z$ are presented in Fig.~5. 
\begin{figure}[h!]
\centering
\includegraphics[angle=0,width=1.0\linewidth]{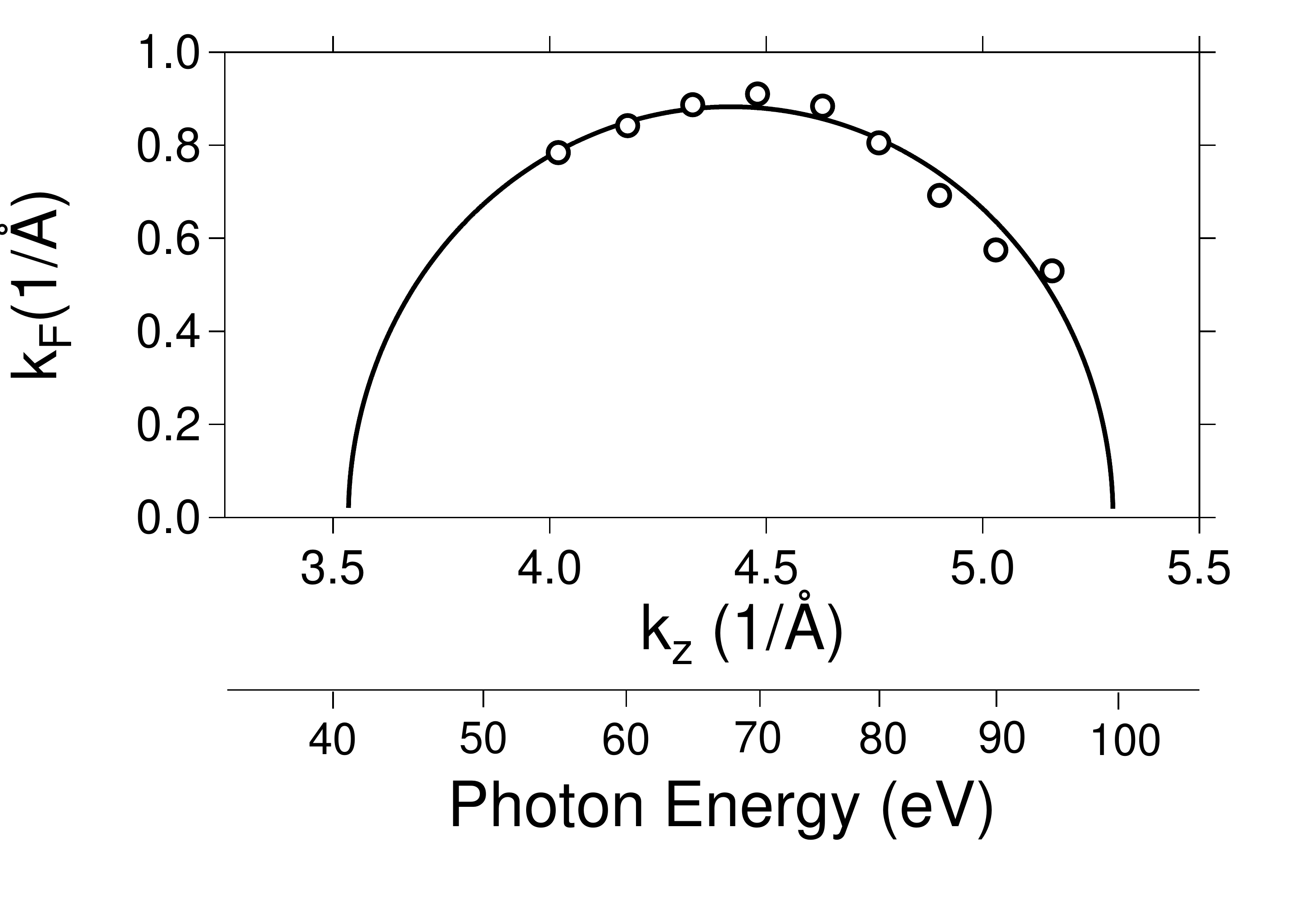}
 \caption
{Fermi wave vector $k_{\mathrm{F}}$ (black markers) of sodium as a function of the photoelectron wave vector $k_z$, normal to the sample surface, or the photon energy. Black solid line: least square fit of the data to a circular Fermi surface.
}
\centering
\end{figure}
We have added a  scale of  the photon energy. The experimental data were fitted by
\begin{equation}
k_F(k_z)=(\overline{k}_F^2-(k_z-k_{z0})^2)^{0.5}.
\end{equation}

We obtain for the average  Fermi wave vector $\overline{k}_F= $0.88 (1/\AA), which is close to the value $k_F=$ 0.91 (1/\AA) derived from the $k_{||}=k_y$ dependent dispersion presented in the main paper.
The maximum of $k_F(k_z)$ appears at $k_z=$4.48 (1/\AA) or $h\nu=$69 eV . This shows that the  photon energy $h\nu=$70 eV used in the main paper is close to a value corresponding  to a  plane with $k_z =$0. The results demonstrate the  nearly-free 3D nature of the Fermi surface.

\section{\label{sec:AB}  CORRECTIONS FOR ELASTIC SCATTERING AND FINAL STATE EFFECTS}
The experimental data on scattering rates $\Gamma_{\mathrm{exp}}$ are composed out of three contributions: the inelastic scattering rate of the photohole $\Gamma^{\mathrm{in}}_{\mathrm{h}}$, which is the central quantity of the main paper, the elastic scattering rate of the photohole $\Gamma^{\mathrm{el}}_{\mathrm{h}}$ due to impurities, and contributions due to the finite lifetime  broadening of the excited photoelectron in the final state $\Gamma_{\mathrm{e}}$. In a Fermi liquid, at the Fermi surface, $\Gamma^{\mathrm{i}}_{\mathrm{h}}$ is expected to be zero. Then $\Gamma_{\mathrm{exp}}(0)$ is  the sum of $\Gamma^{\mathrm{el}}_{\mathrm{h}}(0)$ and $\Gamma_{\mathrm{e}}(0)$.  If the energy dependence of the  latter two quantities is known, the experimental scattering rates can be corrected in the entire energy range to yield $\Gamma^{\mathrm{i}}_{\mathrm{h}}(E)$~\cite{Damascelli2003,Pendry1978,Eastman1978,Thiry1979,Knapp1979,Chiang1980,Kevan1980,Grepstad1982,Bartynski1986,Smith1993,McLean1994}.
\par
In Fig.~6 we plot the total lifetime broadening derived from the ARPES data shown in Fig.~3 of the main paper. As described above,
the finite offset at the Fermi level is due to the elastic scattering $\Gamma^{\mathrm{el}}_{\mathrm{h}}(0)$ and/or due to $\Gamma_{\mathrm{e}}(0)$.
\par
First, we discuss the case where the offset is completely determined by elastic scattering.
Usually, it is assumed that the elastic scattering rate $\Gamma^{\mathrm{el}}_{\mathrm{h}}$
is independent of the energy $E$. As it is caused by 
elastic scattering by impurities which are separated by
a mean distance $d$, the inverse lifetime $1/\tau^{\mathrm{el}} = \Gamma^{\mathrm{el}}_{\mathrm{h}} =
dv_{\mathrm{h}}$, where $v_{\mathrm{h}}$ is the group velocity of the quasiparticles/photoholes.
Close to the Fermi level, it is a good approximation to
assume a linear dispersion from which follows that $v_{h}$ is
constant and therefore $\Gamma^{\mathrm{el}}_{\mathrm{h}}$ should be constant. On the
other hand, in the present work, we study the spectral
weight over a large energy range and the dispersion is
parabolic leading to an energy dependent velocity. Close
to the Fermi level it is high while at the bottom of the
band it is zero. Thus, the elastic scattering rate should
be $\Gamma^{\mathrm{el}}_{\mathrm{h}}(E) = (v_{\mathrm{h}}(E)/v_{\mathrm{h}}(0))\Gamma^{\mathrm{el}}_{\mathrm{h}}(0)$.
The corrected values $\Gamma^{\mathrm{in}}_{\mathrm{h}}(E)=\Gamma_{\mathrm{exp}}(E)-\Gamma^{\mathrm{el}}_{\mathrm{h}}(E)$ are presented in Fig.~6 by green markers. From a least squares fit with   $ \Gamma^{\mathrm{in}}_{\mathrm{h}}= \alpha E^n$, we obtain the parameters $\alpha=0.118 \pm 0.002$ and $n=2.06 \pm 0.02$.
   
\par
Next, we discuss the case where the offset of the width at the Fermi level is completely determined by 
a finite lifetime broadening of the photoelectrons. Following the register-line formalism~\cite{Himpsel1980,Fraxedas1990,McLean1994}, the experimental broadening $\Gamma_{\mathrm{exp}}$ is given by
\begin{equation}\label{eq:GTFE}
\Gamma_{\mathrm{exp}} =(\Gamma_{\mathrm{h}}+R\Gamma_{\mathrm{e}})/(1-R).
\end{equation}
where $\Gamma_{\mathrm{h}}$ and $\Gamma_{\mathrm{e}}$ are the inverse lifetime of the photohole and the photoelectron, respectively. R is given by
\begin{equation}\label{eq:RGRAD}
R=\frac{{\bf \hat e} \cdot {\bf\nabla_k} E_{\mathrm{h}}}{\bf {\hat e} \cdot \bf{\nabla_k} E_{\mathrm{e}}}.
\end{equation}
${\bf \hat e}$ is a unit vector which is tangential to the RL, and ${\bf \hat y}$ and ${\bf \hat z}$
are unit vectors determining the  $y-z$ scattering plane.

We approximate the dispersion of the photohole and the photoelectron  by a free-electron band  $E=(\hbar^2/2m)(k_y^2+k_z^2)$. Then the gradients in Eq.~\ref{eq:RGRAD} are
\begin{equation}\label{eq:GRAD}
{\bf\nabla_k}E(k)=(\hbar^2/m)({\bf \hat y}k_y+{\bf \hat z}k_z).
\end{equation}
For a free photoelectron state the RL is given by
\begin{equation}\label{eq:REGL}
k_z=(\frac{2m_0}{\hbar^2}V_0+k_y^2\cot^2\Theta)^\frac{1}{2}.
\end{equation}\label{eq:ERG}
$\Theta$ is the scattering angle relative to ${\bf \hat z}$. The inner potential $V_0=$ 10  eV was derived from the energy dependent ARPES experiment on Na~\cite{Jensen1985}. It is consistent with our data presented in Fig.~5 of Appendix A.
The unit vector ${\bf \hat e}$ is determined by
\begin{equation}\label{eq:ERG}
{\bf \hat e}={\bf \hat y}e_y+{\bf \hat z}e_z=\frac{{\bf \hat y}+{\bf \hat z}(k_y/k_z) \cot^2\Theta}{((k_y/k_z)^2\cot^4\Theta+1)^\frac{1}{2}}
\end{equation}
In the present experiment, we measure the photohole dispersion for $k_z=$0. Then combining Eqs.~\ref{eq:GRAD} and \ref{eq:ERG} and using a renormalized free-electron band with an effective mass $m^*$, we obtain
\begin{equation}\label{eq:RFIN}
R=\sin^2\Theta/m^*
\end{equation}

\begin{figure}[t!]
\centering
\includegraphics[angle=0,width=1.0\linewidth]{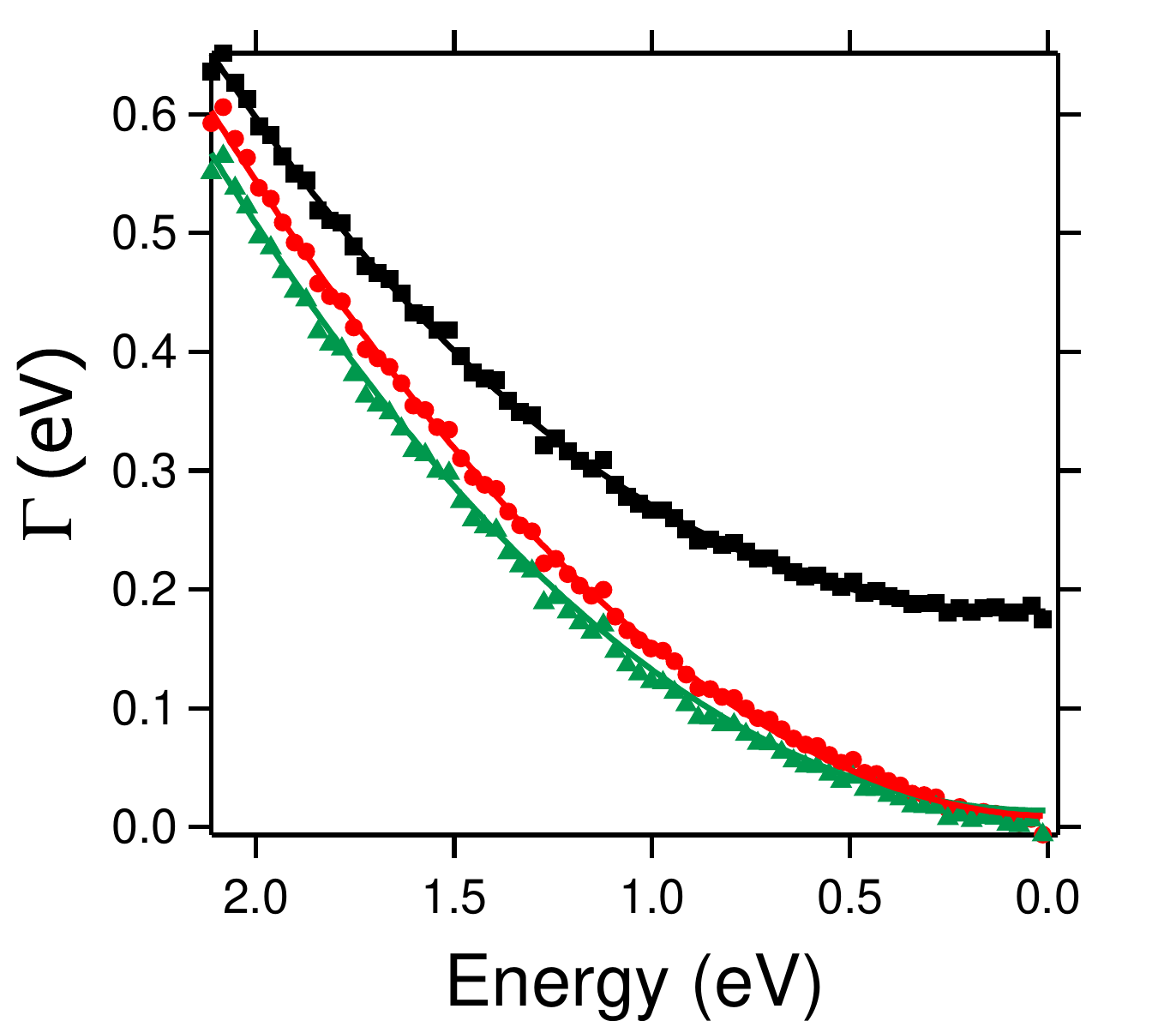}
 \caption
{Presentation of the total scattering rates derived from ARPES (black markers) together with data corrected for energy dependent elastic scattering $\Gamma_{\mathrm{h}}^{\mathrm{el}}(E)$ (green markers) and data corrected for energy dependent ``final state`` broadening $\Gamma_{\mathrm{h}}^{\mathrm{i}}(E)$(red markers). The solid lines are least squares fit with $\Gamma_{\mathrm{h}}^{\mathrm{i}}(E)= \alpha E^n$.
} 
\centering
\end{figure}
\par
For a Fermi liquid the photohole broadening $\Gamma_{\mathrm{h}}(0)$ at the Fermi level is zero and assuming that the elastic scattering broadening is negligible, the experimental broadening $\Gamma_{\mathrm{exp}}(E)$ is completely caused by the lifetime broadening of the photoelectron, which can be calculated from
\begin{equation}\label{eq:NOM} 
\Gamma_{\mathrm{e}} =\Gamma_{\mathrm{exp}}(0)(1-R(0))/R(0).
\end{equation}
On the other hand, if  the final state broadening $\Gamma_{\mathrm{e}}$ is known, the photohole broadening $\Gamma_{\mathrm{h}}$ can be derived from the experimental data $\Gamma_{\mathrm{exp}}$. This has been discussed and demonstrated in numerous papers in the literature~\cite{Pendry1978,Eastman1978,Thiry1979,Knapp1979,Chiang1980,Kevan1980,Grepstad1982,Bartynski1986,Smith1993,McLean1994}.
\par
Using the experimental broadening at the Fermi level $\Gamma_{\mathrm{exp}}=$0.18 eV, we derive for the photohole broadening 
$\Gamma_{\mathrm{e}}=$4.8 eV. This is a maximal value which may be reduced by finite contributions from elastic scattering. The value, which is expected to be weakly photon energy dependent, is close to $\Gamma_{\mathrm{e}}\approx$3 eV estimated for a photon energy of 100 eV  derived from an analysis of LEED data~\cite{Pendry1978}.
\par
Here we remark that the finite energy resolution, although considered in our evaluation, has no influence on our results because it is much smaller ($\Delta E=$ 0.03 eV) than the corrections due to elastic scattering and final state effects (0.18 eV). Moreover, we emphasize that an energy dependent matrix element was taken into account in our linewidth analysis~\cite{Kurleto2021}.
\par
Going back to the case of a broadening which is completely caused by ``final state`` effects we derive for the energy dependent initial state broadening
\begin{equation}\label{NOM} 
\Gamma_{\mathrm{h}}^{\mathrm{fs}}(E) =\Gamma_{\mathrm{exp}}(E)(1-R(E))-R(E)\Gamma_{\mathrm{e}}
\end{equation}
The data for $\Gamma_{\mathrm{h}}^{\mathrm{fs}}(E)$ corrected for the finite lifetime of the photoelectron  are depicted in Fig.~6 by red markers. In this case a least squares  fit yields $\alpha=0.143 \pm 0.002$ and $n=1.90\pm 0.02$. The parameters and the corrections are not very different
from those using solely  elastic scattering  contributions.
\par
This comparison shows, that the parameters describing  the energy dependent $e-e$ scattering rates $\Gamma_{\mathrm{h}}$, presented in the main paper, are nearly independent of the character of the corrections.
In the main paper, we present in Fig.~4 the mean value for $ \Gamma_{\mathrm{h}}=(\Gamma_{\mathrm{h}}^{\mathrm{el}}+\Gamma_{\mathrm{h}}^{\mathrm{fs}})/2$ and the corresponding parameters.

\bibliographystyle{apsrev4-2}
\bibliography{Sodium}

\end{document}